\documentclass[twocolumn,prl,showpacs]{revtex4}

\usepackage{graphicx}
\usepackage{dcolumn}
\usepackage{float}
\usepackage{amsmath}

\newcommand{\comment}[1]{}

\begin{document}

\title{\boldmath High-transition-temperature superconductivity \\
in the absence of the magnetic-resonance mode\unboldmath}

\author{J. Hwang$^{1}$, T. Timusk$^{1}$ and G.D. Gu$^{2}$
} \email{timusk@mcmaster.ca}

\affiliation{$^{1}$Department of Physics and Astronomy, McMaster
University, Hamilton, ON L8S 4M1, Canada\\ $^{2}$Department of
Physics, Brookhaven National Laboratory, Upton, New York 11973-5000,
USA } \date{\today}


\maketitle


{\bf The fundamental mechanism that gives rise to
high-transition-temperature (high-$T_c$) superconductivity in the
copper oxide materials has been debated since the discovery of the
phenomenon. Recent work has focussed on a sharp 'kink' in the
kinetic energy spectra of the electrons as a possible signature of
the force that creates the superconducting
state~\cite{rossat-mignod91,mook98,fong99,he01,kaminski01,
johnson01,norman98,campuzano99,abanov99,zasadzinski01,thomas88,puchkov96,
carbotte99,munzar99}. The kink has been related to a magnetic
resonance~\cite{carbotte99,dai99,demler98,scalapino99} and also to
phonons ~\cite{lanzara01}.  Here we report that infrared spectra
of Bi$_{2}$Sr$_{2}$CaCu$_{2}$O$_{8+\delta}$ (Bi-2212) show that
this sharp feature can be separated from a broad background and,
interestingly, weakens with doping before disappearing completely
at a critical doping level of 0.23 holes per copper atom.
Superconductivity is still strong in terms of the transition
temperature ($T_{c}\approx$ 55 K), so our results rule out both
the magnetic resonance peak and phonons as the principal cause of
high-$T_c$ superconductivity. The broad background, on the other
hand, is a universal property of the copper oxygen plane and a
good candidate for the 'glue' that binds the electrons.}


We investigated the Bi-2212 material systematically as a function
of doping, including highly overdoped Bi-2212 (see Methods). To
better display the sharp 'kink', we do not show the optical
conductivity but focus our attention on a related quantity, the
optical single particle self-energy, $\Sigma^{op}(\omega)$ (see
Methods). In Fig. 1{\sf a}-{\sf d} we plot the imaginary part of
this quantity which is simply the scattering rate of the charge
carriers. At high frequencies the scattering rate varies linearly
with frequency: this is called the marginal Fermi liquid (MFL)
behavior~\cite{varma89}. We note that the overall scattering rate
decreases as the doping increases and that there is a sharp
depression in 1/$\tau(\omega)$ below 700 cm$^{-1}$ at low
temperatures, with an overshoot in the superconducting state. This
sharp onset of scattering has been in general attributed to the
interaction of the charge carriers with a bosonic mode and more
recently to the 41 meV neutron
resonance~\cite{norman98,carbotte99,munzar99}. In the real part of
the optical self-energy, plotted in panels 1{\sf e}-{\sf h}, we
note a sharp peak around 700 cm$^{-1}$, which tracks the
depressions in 1/$\tau(\omega)$ in both frequency and amplitude.
One very interesting property of the peak in the self-energy is
its doping and temperature dependence: the peak gets weaker as the
doping level and  the temperature increase. We will call the peak
the 'optical resonance mode'. The resonance peak in the
self-energy spectrum can clearly be resolved from the broader
background of MFL scattering.
\begin{figure}[t]
  \vspace*{-0.4cm}%
  \centerline{\includegraphics[width=3.5 in]{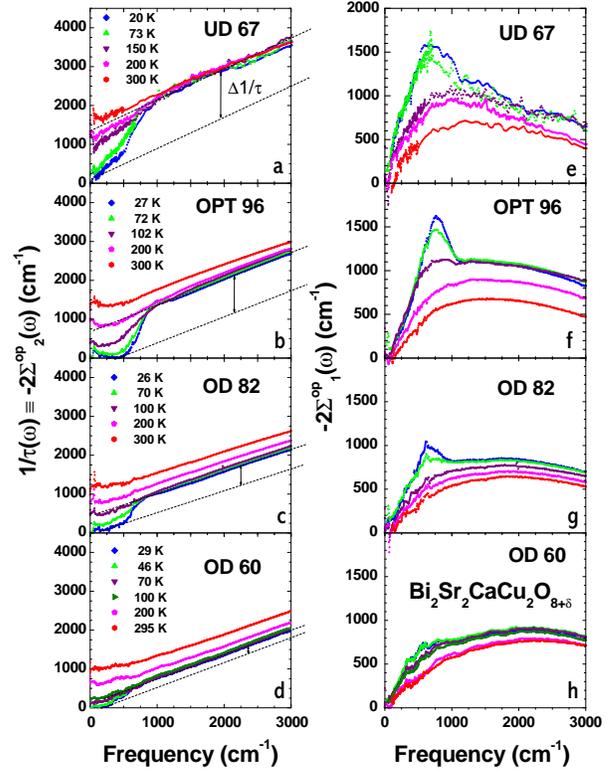}}%
  \vspace*{-1.0cm}%
  \caption{The optical single-particle self-energy of
Bi$_{2}$Sr$_{2}$CaCu$_{2}$O$_{8+{\delta}}$. {\bf {\sf a}}-{\bf
{\sf d}}, The doping and temperature dependent optical scattering
rate, $1/\tau(\omega)$ for four representative doping levels.
{\bf{\sf a}}, $T_{c}=67$ K (underdoped); {\bf{\sf b}}, 96 K
(optimal); {\bf{\sf ac}}, 82 K (overdoped); {\bf{\sf d}}, 60 K
(overdoped). {\bf {\sf e}}-{\bf {\sf h}}, The real part of the
optical self-energy, $-2\Sigma^{op}_{1}(\omega)$. The slope of
$-\Sigma^{op}_{1}(\omega)$ near $\omega=0$, is closely related to
the coupling constant or the mass enhancement factor, and also
decreases as the doping increases. This is consistent with other
studies~\cite{hwang03,johnson01}. We note the weakening of the
feature at 700 cm$^{-1}$ in both sets of curves as the doping
level increases.}
  \label{Fig1}
\end{figure}

\begin{figure}[t]
  \vspace*{-0.4cm}%
  \centerline{\includegraphics[width=3.5 in]{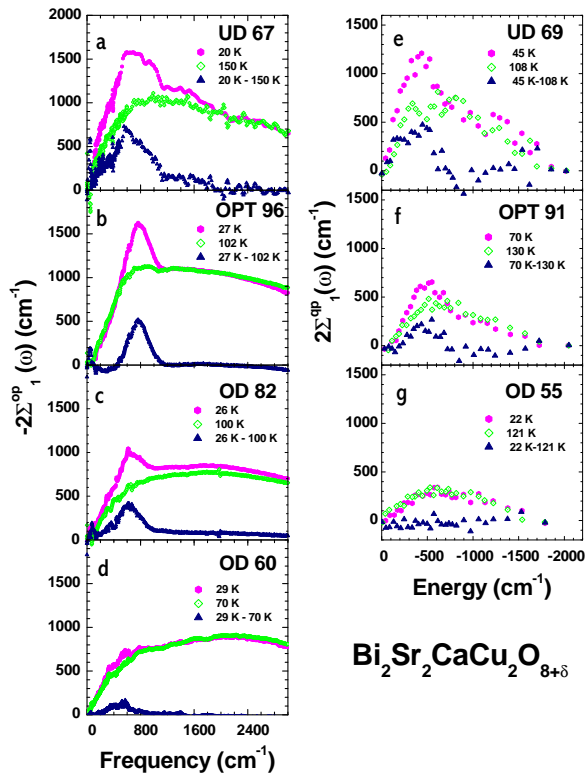}}%
  \vspace*{-1.0cm}%
  \caption{Comparison of the self energy measured with infrared and
angle-resolved photoemission for Bi-2212. {\bf {\sf a}}-{\bf {\sf
d}}, The real part of the optical self-energies,
$-2\Sigma^{op}_{1}(\omega)$ of the normal and superconducting
phases for a series of crystals with different doping levels
determined by FTIR spectroscopy, as well as the difference between
normal and superconducting state curves. {\bf{\sf a}}, $T_{c}=67$
K (underdoped); {\bf{\sf b}}, 96 K (optimal); {\bf{\sf ac}}, 82 K
(overdoped); {\bf{\sf d}}, 60 K (overdoped). {\bf {\sf e}}-{\bf
{\sf g}}, The real part of the self-energies,
$-2\Sigma^{qp}_{1}(\omega)$ from ARPES measurements of ref. 6.
{\bf{\sf e}}, $T_c=69$ K (underdoped); {\bf{\sf f}}, $T_c=91$ K
(optimal); {\bf{\sf g}}, $T_c=55$ K (overdoped). Although the
absolute magnitudes and frequencies differ for the two data sets,
the overall qualitative features are closely correlated when one
allows for the higher noise level and lower resolution of the
photoemission data.}
  \label{Fig2}
\end{figure}

The optical self-energy is closely related to the quasiparticle
self-energy $\Sigma^{qp}(\omega)$, which can be measured in ARPES
experiments although there are important differences in the two
quantities~\cite{kaminski00,schachinger03,millis03}. In Fig. 2 we
show a comparison of the optical self-energy from our data and the
ARPES self-energy at two temperatures: one in the normal state and
the other the superconducting state. We see that although the
qualitative features are very similar, the magnitudes and the
frequency values of the various features differ considerably, as
expected from theory. Part of the difference can be attributed to
the uncertainty in locating the unrenormalized base line in the
ARPES analysis. As noted in ref. 6, a peak in
$\Sigma^{qp}_{1}(\omega)$ can be clearly separated from the broad
continuum and is closely correlated with the neutron resonance
mode. The lower noise level and the higher resolution of our
optical data demonstrate a clear doping dependent trend -- the
amplitude of the resonance peak weakens with doping in the
overdoped region and disappears completely at a doping level where
the superconductivity is still large enough to show $T_{c}\approx
55$ K, as shown in Fig. 2{\sf g}. Thus, at this doping level we
have superconductivity without the resonance peak, and whereas the
peak may contribute to the condensation
energy~\cite{demler98,dai99a} and the superconducting gap at lower
doping levels, it cannot be the main cause of high-$T_c$
superconductivity.

In Fig. 3 we investigate further the disappearance of the optical
resonance mode as a function of doping. In Fig. 3{\sf a} we plot
the amplitude of the resonance mode in the real part of the
optical self-energy, which decreases uniformly with doping and
appears to extrapolate to zero at a doping level $p=$ 0.23
($T_{c}\approx 55$ K). The center frequency of the mode (Fig.
3{\sf b}) is proportional to the transition temperature
($\Omega^{op}_{res} \approx 8.0 k_BT_c$) and reaches a maximum at
the optimally doped phase for both FTIR and ARPES, which is
consistent with other studies~\cite{he01}. In Fig. 3{\sf c} we
show the contribution of the resonance mode to the imaginary part
of the self-energy as a function of doping, both directly and
normalized as a percentage of the total scattering rate at 3000
cm$^{-1}$ (see Methods). This is done to divide out the overall
decrease of the coupling of the charge carriers to the bosonic
fluctuations that occurs with increased
doping~\cite{johnson01,hwang03}.

\begin{figure}[t]
  \vspace*{-0.4cm}%
  \centerline{\includegraphics[width=3.5in]{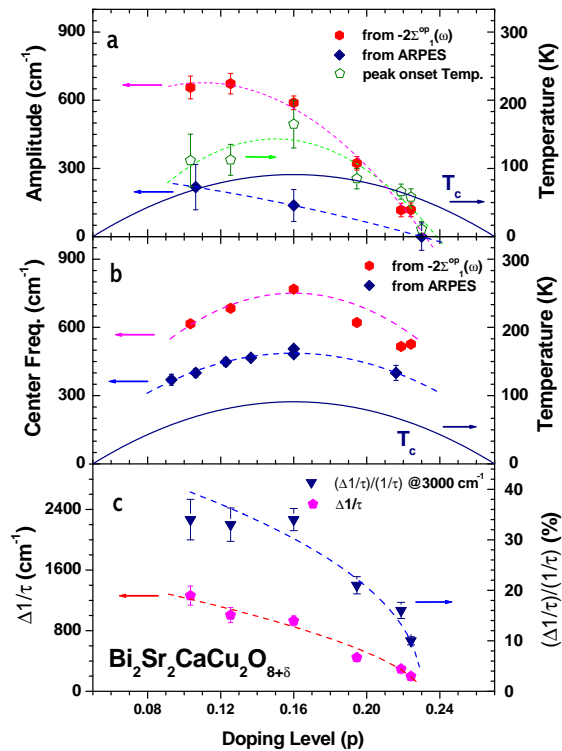}}%
  \vspace*{-1.0cm}%
  \caption{Doping-dependent properties of the optical resonance mode. See the
Methods section for detail descriptions. {\bf {\sf a}},
Amplitudes; {\bf {\sf b}}, center frequencies. The solid curve is
$T_{c}(p)$ and the two dashed curves are parabolic fits, $5.6
k_{B}T_c$ for ARPES and $8.0 k_{B}T_c$ for FTIR. {\bf \mbox{\sf
c}}, the quantity $\Delta 1/\tau$ (see Fig. 1 {\bf {\sf a}}-{\bf
{\sf d}}), the contribution of the resonance mode to the
scattering rate, as a function of doping, and the same quantity
normalized to the scattering rate at 3000 cm$^{-1}$. All the data
point to the disappearance of the mode at a critical doping level
of $p=0.23$.} \label{Fig3}
\end{figure}

We do not have detailed data on the amplitude of the mode as a
function of temperature. In the region of $p=0.19$ the mode
appears at the superconducting transition temperature of 80 K
along with the neutron mode~\cite{fong99,he01}. The error in the
critical doping value cannot be found from optical data alone
because we were not able to obtain a sample with sufficient
overdoping. The ARPES data of ref. 6 for their $T_c=55$ K sample
shows no resonance, so we can use this as an upper limit and our
own 60 K sample as a lower limit, giving $p=0.225 \pm 0.010$ for
the value of the critical doping point.

While it is clear from our results that the amplitude of the
coupling of the optical resonance mode to charge carriers
extrapolates to zero at a critical doping of $p=0.23$, we are not
able to determine the behavior of the center frequency of the
resonance mode at the critical doping point. The center frequency
appears to drop below the parabolic trend, shown as the dashed
curve in Fig. 3{\sf b}. According to theory, the frequency of the
ARPES resonance ($\Omega^{qp}_{res}$) is roughly the sum of the
gap ($\Delta$) and the frequency of the neutron mode ($\Omega$)
\cite{carbotte99} (we note that in optics $\Omega^{op}_{res}=
2\Delta + \Omega$).

We remark here on the relationship between our work and the
proposal of Lanzara {\it et al.}~\cite{lanzara01} that the sharp
kink seen in angle-resolved photoemission is due to phonons. With
our higher resolution and lower noise level we are able to
separate the sharp peak from the overall broad background. The
broad background can be seen in the spectra of all high-$T_c$
superconductors and seems to represent a universal property of the
copper oxide plane. But it cannot be due to phonons, because its
spectral weight extends beyond the cut-off frequency of the phonon
spectrum~\cite{schachinger03}. The sharp structure, on the other
hand, has all the characteristics of the magnetic resonance: it
makes its appearance at the superconducting $T_{c}$ in overdoped
copper oxides and slightly above $T_{c}$ in the underdoped copper
oxides. We have also shown that it vanishes completely at a high
doping level inside the superconducting dome. None of this is
expected for phonons which should be present at all temperatures
and doping levels.

The disappearance of the resonance mode may be related to the
possible vanishing of the pseudogap and the existence of a quantum
critical point~\cite{chakravarty01,loram00}. Recent
data~\cite{shibauchi01} suggests that the critical doping level
for the disappearance of the pseudogap in c-axis transport may be
higher than $p=0.22$ at least in Bi-2212. The pseudogap has less
influence than the resonance mode on ab-plane properties as shown
by work on the anisotropy of dc transport~\cite{takenaka94}. Also
on theoretical grounds, a gap in the fluctuation spectrum will not
give the upward step in the scattering rate that we
observe~\cite{norman98}.  The disappearance of the resonance mode
has been predicted to occur near the $p=0.22$ doping level within
the magnetic exciton picture~\cite{abanov99,zasadzinski01}. Our
analysis of the optical self-energy, which allows us to isolate
the resonance peak from the universal broad background in the
single-particle self-energy for Bi-2212, should work for other
copper oxides and gives us a direct tool for examining the
spectrum of excitations responsible for pairing in high-$T_c$
superconductors.

$\\$

{\bf Methods}

{\bf Experiments and sample preparations}

We used Fourier-transform infrared (FTIR) spectroscopy to obtain
the reflectance data~\cite{homes93} of floating-zone-grown
single-twinned crystals of Bi-2212. The highly overdoped samples
were made by annealing the as-grown crystals at high temperature,
$T > 500$ $^{\circ}$C, in 3000 bar liquid oxygen in sealed
containers. We present new data on one optimally doped Bi-2212
($T_{c}=96$ K), and three highly overdoped samples ($T_{c}=82$ K,
$T_{c}=65$ K, and $T_{c}=60$ K). The four new samples are bulk
superconductors with London penetration depths of 2,065 $\AA$,
1,638 $\AA$, 1,481 $\AA$ and 1,375 $\AA$, respectively. The
optimally doped sample has been doped with a small amount of Y to
yield a relatively well-ordered system~\cite{eisaki03}. To cover a
wide range of doping levels we also used data from previously
measured underdoped Bi-2212 crystals~\cite{puchkov96}.

{\bf The optical single particle self-energy,
$\Sigma^{op}(\omega)$}

The optical conductivity was determined by Kramers-Kronig
analysis~\cite{homes93} and we present our data within the
framework of the extended Drude model, where the complex optical
conductivity can be written as $\sigma(\omega)= -i
\omega(\epsilon(\omega) - \epsilon_H)/4\pi= -i
\omega_{p}^{2}/[4\pi(2\Sigma^{op}(\omega)-\omega)]$, where
$\epsilon_H$ is the dielectric constant at a high frequency ($\sim
2 eV$) for each doping level (note that there is doping dependence
in $\epsilon_H$), $\omega_{p}$ is the plasma frequency and
$\Sigma^{op}(\omega)=\Sigma^{op}_{1}(\omega)+i\Sigma^{op}_{2}(\omega)$
is the optical single-particle self-energy. In terms of more
familiar quantities $\Sigma^{op}_{1}(\omega)\equiv
\omega(1-m^*/m)/2$, where $m^*/m$ is the mass renormalization, and
$\Sigma^{op}_{2}(\omega)\equiv -1/(2\tau(\omega))$, where $\tau$
is the carrier lifetime. We determine the doping-dependent plasma
frequency from the absorption spectra in the near-infrared
region~\cite{hwang03}.

{\bf Doping dependence of the optical resonance mode in
$\Sigma^{op}(\omega)$}

We used the parabolic approximation to obtain the doping levels
from the transition temperature, $T_{c}$, with the slope of the
infrared reflectance as an additional test of the doping
level~\cite{hwang03}. The amplitude and center frequency of the
optical resonance mode are obtained from
$-2\Sigma^{op}_{1}(\omega)$ data by a least-squares fit of the
peak to a Lorentzian line shape (see Fig. 3{\bf {\sf a}} and {\bf
{\sf b}}). To estimate the strength of the interaction of the
resonance mode with the carriers from
$1/\tau(\omega)\equiv-2\Sigma^{op}_{2}(\omega)$, we draw a dashed
line, parallel to the high frequency trend, from the onset point
of substantial scattering. The difference between this line and
the actual high frequency scattering, $\Delta 1/\tau$ is our
estimate of the contribution of the sharp mode to the scattering.
It exhibits a doping dependence that is suggestive of mean-field
behavior (dashed curve) with a critical point at $p=0.23$. (see
Fig. 1{\bf {\sf a}}-{\bf {\sf d}} and Fig. 3{\bf {\sf c}}). We
estimate the doping dependence of the onset temperature of the
resonance mode from both FTIR and ARPES data. We have drawn the
points as follows: the upper error-limit temperature where there
is no resonance. The lower-limit temperature is where we first see
the resonance. The resonance onset temperature is the average of
the upper and lower limits (see Fig.  3{\sf a}).

{\bf Acknowledgments}

This work has been supported by the Canadian Natural Science and
Engineering Research Council and the Canadian Institute of Advanced
Research. We thank H. Eisaki and M. Greven for supplying us with
several crystals. Their work at Stanford University was supported by
the Department of Energy's Office of Basic Sciences, Division of
Materials Science.  The work at Brookhaven was supported in part by
the the Department of Energy. We thank D.N. Basov, J.P. Carbotte,
G.M.~Luke and M.R. Norman for helpful discussions.

\end{document}